# Million Degree Plasmas in Extreme Ultraviolet (EUV) Astrophysics

A Science White Paper submitted to the Astro 2010 Decadal Survey

Lead Author: Michael P. Kowalski, NRL, 202-767-2781, michael.kowalski@nrl.navy.mil

Co-Authors:  Martin Barstow, UL
Frederick Bruhweiler, CUA
Raymond Cruddace, NRL
Andrea Dupree, SAO
Jay Holberg, UA/LPL
Steve Howell, NOAO
J. Martin Laming, NRL
Jeffrey Linsky, JILA
Edward Sion, VU
Tod Strohmayer, NASA-GSFC
Paula Szkody, UW
Barry Welsh, UCB
Michael Wolff, NRL
Kent Wood, NRL

**Abstract:**

Million degree plasmas are ubiquitous in the Universe, and examples include the atmospheres of white dwarfs; accretion phenomena in young stars, cataclysmic variables and active galactic nuclei; the coronae of stars; and the interstellar medium of our own galaxy and of others. The bulk of radiation from million degree plasmas is emitted at extreme ultraviolet (EUV) wavelengths, which includes critical spectral features containing diagnostic information often not available at other wavelengths. With underpinning by a mature instrument technology, there is great opportunity here for exciting discoveries.

**I. Introduction**

Million degree plasmas are ubiquitous in the Universe, and examples include the atmospheres of white dwarfs (WDs); accretion phenomena in young stars (classical T-Tauri), cataclysmic variables (CVs) and active galactic nuclei (AGN); the coronae of stars; and the interstellar medium (ISM) of our own galaxy and of others. Understanding their nature is fundamental to astrophysics. This White Paper addresses primarily two Astro2010 Thematic Science Areas, *SSE* and *GAN*, and has important implications for the other three areas.

The extreme ultraviolet (EUV:~90-912 Å) includes critical spectral features containing diagnostic information often not available at other wavelengths (e.g., He II Ly series 228-304 Å, Fe IX-XXIV), and the bulk of radiation from million degree plasmas is emitted in the EUV (e.g., Fig. 2 http://chips.ssl.berkeley.edu/science.html#sec1). However, sensitive EUV spectroscopy of high resolving power ($R=\lambda/\Delta\lambda=3{,}000\text{-}10{,}000$) is required to resolve source spectral lines and edges unambiguously, to identify features produced by the intervening ISM, and to measure line profiles and Doppler shifts. This allows exploitation of the full range of plasma diagnostic techniques developed in laboratory and solar physics, where ongoing solar EUV observations (e.g., flares, CMEs) invite fruitful comparisons to stars. The combination of high-resolution spectroscopy with sensitive photometry, timing measurements of suitable cadence, and advanced theory are the keys to understanding the physics of million degree plasmas in these objects.

Recent technical advances (multilayer-coated optics) have produced the first astrophysical high-resolution EUV spectrometer, *J-PEX*, which has made two successful sounding rocket flights, during which spectra were obtained for the WDs G191-B2B (Fig. 1) and Feige 24. These were combined with a spectral modeling code capable of managing millions of lines (TLUSTY: http://nova.astro.umd.edu/) to provide significant scientific results for G191-B2B.[1-2] Analysis of the Feige 24 data is proceeding. Future opportunities are compelling; here we assume a SMEX (sec. V). We call out four specific but compound questions that are ripe for answering and a general area with great discovery potential (beyond SMEX):

1. What is the evolutionary history of WDs?
2. What are the densities, temperatures, ionization states, and depletion levels of the local ISM (LISM)?
3. Where are the emission sites in CVs and what are their temperatures, densities, compositions, and dynamics?
4. What are the structure, dynamics, and evolution of stellar coronae? What is the structure of accretion columns and shocks in young stars? What is the relationship of coronal heating and flares? What is the relative importance of magnetic reconnection and MHD/acoustic waves in heating? How do coronal abundances differ from those in the photosphere? How do stellar coronae influence planetary atmospheres?

Discovery Area: As observations are pushed to cosmological distances ($z>5$) the spectral energy distributions of bright X-ray objects, AGNs for example, will have their maxima redshifted into the EUV, where patches in the LISM at high galactic latitude becomes transparent enough to allow extragalactic observations at its shortest wavelengths, thus providing unprecedented opportunities to observe evolution. (The Lockman hole, $N(H\ I)\sim 4\times 10^{19}$ cm$^{-2}$, is approximately unit optical depth at ~120 Å, and transparency increases rapidly towards shorter wavelengths.)

**II. White Dwarfs, CVs, and the ISM**
**WDs as Compact Objects.** WDs were the first compact, degenerate astronomical objects discovered. Connections exist between WDs and neutron stars (NSs). For example, dwarf novae and classical novae, where the WD is accreting from a companion, are mirrored by low/high mass X-ray binaries and bursters, in which a NS is the compact object. WD masses lie in the range 0.2-1.4 $M_\odot$, while NSs are formed at and remain near 1.4 $M_\odot$. Spectra of hot WDs peak in the EUV where most atomic transitions lie, providing diagnostics not generally available in NSs. Models with scaled parameters reinforced by spectral and timing measurements allow fruitful comparisons of analogous systems. Source class characterization is mature such that analogies can be used to compare physical processes across a wide dynamic range of parameters, one example being theories of quasiperiodic oscillations.[3] WDs also represent key laboratories for studying the physics of stellar atmospheres in high gravitational fields.

**WDs and Stellar Evolution.** White dwarfs are among the oldest stars and are the end products of most stars in the galaxy. Their distributions help determine disk age and map the history of Galactic star formation. Yet white dwarf evolution is complicated and poorly understood. Ascertaining their cooling ages and masses calls for thorough knowledge of how photospheric compositions evolve. Metal abundances affect cooling rates and bias determinations of temperature and surface gravity, which must be known accurately to derive a reliable mass. There are two main types of hot WDs: H-rich DA and He-rich DO and DB stars. Abundances should reflect evolutionary paths; however, observed patterns appear to be affected by gravitational settling (diffusion), selective radiative levitation, possible mass loss, magnetic fields, and accretion, which may include planetary debris.[4] The $T_{eff}$, log g, and the interstellar environment may govern the relative importance of these processes. Understanding of WD evolution is incomplete; for example a dearth of stars in the He-rich cooling sequence at ~30000-45000 K is unexplained, unless these objects temporarily masquerade as H-rich DA WDs. Thus the detection of photospheric He in DA WDs is critical to finding a solution. For low mass stars, the final He abundance is determined by core nucleosynthesis, shell-burning, and stellar mass-loss, and after gravitational settling has further altered photospheric composition, allowing He to sink below the photosphere to leave overlying H. For WDs in binaries, mass loss may be affected by a common envelope phase that leaves a H-rich shell thinner than expected on a single star.

Helium can be detected in the optical/FUV wavebands, but EUV spectroscopy is a hundred times more sensitive. Significant quantities of heavier elements are also present in the hottest (T>50,000 K) WD atmospheres. Lines are formed at varying atmospheric depths in the EUV and FUV, and abundance measurements from high-resolution spectroscopy in both bands are required. FUV data from *IUE* or *HST* are available for many stars, but *EUVE* lacked the spectral resolution to determine abundances uniquely for heavy elements. *Key Objective: Helium in WD Evolution:* Obtaining high-resolution EUV spectra of a diverse WD sample is a key step in understanding WD evolution. A systematic survey for photospheric He should include isolated DA WDs and those in binaries, and cover a range of $T_{eff}$, g, and heavy element compositions. *Key Objective: Heavy Elements in WD Atmospheres:* High-resolution EUV spectra of heavy element-rich DA stars allow measurement of photospheric abundances and their depth dependence. Those DAs with apparently pure H envelopes are an important control group, but improved sensitivity may allow detection of trace elements far below current limits, where present theory predicts. Heavy element transport is critical to models of how thermonuclear releases are initiated in classical novae; they are difficult to trigger without heavy elements. Such transport cannot be studied in NSs, although nuclear releases do occur there.

**WDs in CVs.** CVs encompass various WD binary systems with substantial mass transfer. CVs provide the best laboratory for accretion processes, involving composition, geometry, and the interaction of plasma with magnetic fields. Accretion may take place via a disk and boundary layer around the WD or, if the WD has a magnetic field strong enough to disrupt the disk, through a stream onto its poles, where stream velocities may reach ~3000 km s$^{-1}$. The magnetic CVs have magnetic moments ($\mu=10^{30}$ -$10^{33}$ G cm$^3$) exceeding those found in any *accreting* NS. In classical novae CVs, accretion of H-rich material leads to thermonuclear runaways on the WD surface and ejects processed material into the ISM. Several distinct regions may produce or modify EUV radiation. Hot gas, primarily from the donor star, is responsible for the high-energy radiation. In non-magnetic systems, the inner disk and boundary layer are the main EUV sources. In magnetic systems the gas streaming onto the WD poles forms a column and passes through a shock near the WD surface. In this case EUV emission arises in the shock or the irradiated WD photosphere. In some systems, material may be accreted at a rate allowing steady nuclear burning, providing yet another component. If it is hot enough, the WD could itself produce EUV radiation, although this shorter wavelength component will be attenuated severely by any heavy elements accumulated from the companion. Disentangling these components requires time- and phase-resolved spectroscopic observations of a sample that spans a range of mass accretion rates.

*EUVE* produced low resolution EUV spectra for each of the main types of CVs, but no clear conclusions about underlying physical processes. The magnetic polars have smooth blackbody continua (10-25 eV), but fits cannot distinguish between models of absorbed blackbodies, stellar atmospheres of solar composition, or power laws. A few spectral features were tentatively identified, but not all. The outburst spectra of dwarf novae (e.g., VW Hyi, U Gem, SS Cyg, and OY Car) show great variety and have prominent yet unidentified features, which might be broadened emission features, blends of many narrow lines, or a continuum with strong, broad absorption features. *Key Objective:* Obtaining time- and phase-resolved high-resolution EUV spectra of a sample of all CV types to determine the nature of the emission for the distinct regions in these systems. Radial velocity measurements of isolated narrow spectral features will determine gas stream motions, accretion velocities, and, of particular importance, robust dynamical WD mass estimates. Mass estimates impact detailed numerical simulations of nova and accretion flows; mass estimates set the scale of the gravitational field, yet reliable, accurate estimates remain scarce. The effects of accretion rate on the emission processes should be examined. Element abundances should be measured to search for effects of common envelope evolution and nuclear burning on the WD surfaces.

**The LISM.** Simulations indicate that in the present Universe a large, even dominant, fraction of baryons could be "hidden" in intracluster and intercluster gas, in a Warm-Hot Ionized Medium (WHIM) at $10^5$-$10^7$ K, which enshrouds clusters of galaxies and permeates galactic disks. Physical processes in the ISM have important implications on how the WHIM is produced and maintained. Hot ambient gas largely filling the LISM may be an extension of the WHIM. The LISM seems to have a highly ionized plasma substrate at T~$10^{6.1}$ K and n~$10^{-2}$ cm$^{-3}$, and has a morphology largely shaped by stellar winds and SN activity, either from the nearby Sco-Cen OB association or older events. Within the Local Void (or Bubble) is a complex of warm (~7,000 K) clouds, in which the Sun is embedded. Hot plasma cools primarily through line emission. Fe is normally the dominant coolant at T >$10^{5.5}$ K, making the Fe abundance a critical measurement.

The LISM (<200 pc) is the best interstellar plasma to study because of low absorption and minimal confusion by overlapping lines and edges from diverse regions. *FUSE* and *Chandra* bring new levels of insight to such studies, and the EUV samples temperatures not accessible to them. Strong EUV continua of hot WDs are ideal backdrops for LISM absorption measurements, which are complementary to emission measurements (e.g., CHIPS) and together constrain the heliospheric, LISM, and intergalactic contributions to the soft X-ray/EUV background.

**Helium in the LISM: Probing Time-Dependent Ionization.** Detection of interstellar He is possible only in the EUV. Surprisingly, He is not neutral. Due to its high ionization potentials, it is insensitive to ambient radiation field. The H ionization ratio is extremely sensitive to the stellar EUV radiation field and reflects photoionization equilibrium. Five WDs with low metallic line blanketing measured by *EUVE* provide the best available data on He ionization in local clouds, indicating a fairly uniform He ionization fraction of ~0.25-0.50. However, He ionization is not completely understood. Calculations show that the observed He ionization cannot be due to photoionization from hot stars or from the surrounding $10^6$ K gas. *EUVE* results imply nearby He is in recombination from an ionizing supernova (SN) event occurring ~2-3 Myr ago, a picture supported by further theoretical analysis and by anomalous $^{60}$Fe deposits in the Earth's crust point at times 2-5 Myr. High resolution (R>3000) EUV spectroscopy is needed to more fully explore the history of the LISM. *Key Objective:* Obtain high-resolution EUV spectra from a diverse set of WDs spanning the He II Lyman series, the He II ionization edge at 228 Å, and the He I autoionization feature at 206 Å to disentangle LISM He II from the WD components. Detection limits ~$10^{16}$ atoms cm$^{-2}$ for both He I and He II should be achievable for even modest S/N (~10:1) spectra. Combining these results with column densities of low ionized species through the local cloud complex (*HST & FUSE*) produces the optimum dataset for testing whether He is in an ionization recombination phase and when the ionization event(s) occurred.

**Physical Conditions of the Hot Gas in the LISM and the Galactic Halo.** What temperature explains the unusual line emission inferred in the soft X-ray background? The best fit is at ~$10^{6.1}$ K. However, the predicted emission line spectrum gives a poor fit to the measured spectral lines for any equilibrium or non-equilibrium ionization model, although part of the problem may lie in inaccurate atomic physics for Si. The long recombination time for this gas also allows one to see signatures of multiple SN events. Another partial explanation is that elements such as Fe are depleted, lengthening cooling time. The hot gas remains buoyant and can rise into the halo and leave the galaxy. This suggests a low Fe abundance and long cooling time for the cosmological WHIM. Finally, charge exchange emission caused by the foreground solar wind impacting ambient gas may account for the bulk of the soft X-ray background. Nearly featureless EUV continua in nearby hot, low metallicity DA WDs offer the best means to detect high ionization absorption features arising in the hot LISM. Models (solar abundances, ionization equilibrium) suggest many potential lines (Ne, Fe, Si) with detectable equivalent widths >0.4 mÅ/100-pc for $10^{5.5}$<T<$10^{6.3}$ K, even with significant Fe depletion (10x), assuming an instrument with reasonable effective area and exposure. *Key Objective:* Determine the temperature, density, and depletion of the hot LISM gas.

### III. Stellar Coronae
While the solar corona has been studied extensively, there is still uncertainty about how it is driven by the convection zone magnetic dynamo and how the dynamo formed and evolved. The H-R diagram has been surveyed at X-ray and EUV wavelengths. Many stars possess coronae hotter than the sun and have diverse characteristics, leading to understanding in the solar case. However, the significance of hot coronae goes beyond stellar evolution. On Earth the development of an atmosphere and hence life has been influenced greatly by the Solar System interplanetary environment, which is shaped by the coronal UV/EUV/X-ray flux, solar wind particles, and mass injections, all of which respond to the solar cycle.[5] EUV observations of activity in those nearby stars in which the Kepler mission discovers Earth-like planets would address their chances for harboring life.

**Very Young Stars.** Young actively accreting pre-Main Sequence stars in nearby associations are excellent candidates for high-resolution EUV spectroscopy. The nearest (50 pc) is TW Hya, which is ~10 Myr old, at the likely epoch of planet formation, and with the Hor (30 Myr) and Tuc (<40 Myr) associations provide a diverse easily EUV-accessible sample. High-resolution EUV spectra will distinguish between coronal emission and that from an accretion stream.

**Main Sequence Stars.** High-resolution EUV spectroscopy of a sample of stars on or near the MS will determine the sources of coronal heating and the structure of the upper transition region and the inner corona. The sample should vary in spectral class and cover a range of rotation period and Rossby number. Heating by magnetic field reconnection in loops may be seen as "nano-flaring," but it is difficult to measure directly. However, heating by waves has observable consequences. Acoustic waves have been ruled out as an effective agent, but MHD (e.g., Alfvén) waves can provide ample flux. The non-thermal (NT) plasma energy density may be obtained from measurements of density and NT broadening of line profiles, which combined with the wave propagation (Alfvén) velocity yields the NT energy flux. Magnetic flux densities can be constrained by measurements of the surface fields. This approach has been applied successfully to solar observations and in studies of ε Eri (STIS). This flux is more than sufficient to heat the corona, but there is evidence not all propagates to the corona, but may lead to mass loss. It is crucial to measure NT line broadening and electron densities at higher temperatures, for stars already observed with STIS or *FUSE*, to assess the role of MHD heating.

**Hot Coronal Plasma in Active Stars.** In many active stars, T and $N_e$ are higher than on the Sun, and often-small high-density regions are located at high latitudes and immersed in an extended corona. The high-T coronal material may be confined, but in rapidly rotating stars may be extended and suffer turbulent broadening. The emission measure distribution of active stars reveal 3 characteristic features: an enhancement at ~$2 \times 10^6$ K from possible solar-like structures, another frequently found around $8 \times 10^6$ K, and finally an elevation sometimes seen at higher T, the sources of which are a mystery, especially the "bump" at $8 \times 10^6$ K, for which $N_e > 10^{12}$ cm$^{-3}$. How is such high-pressure plasma heated and contained on the star? New techniques developed at other wavelengths may be applied to high-resolution EUV spectra. First, optical Zeeman-Doppler imaging has revealed photospheric dark spots on active stars, and traced the magnetic field topology. Doppler imaging using coronal EUV lines is simpler than with photospheric lines because the plasma is optically thin, but its resolution is limited by thermal broadening. Second, phase-resolved Doppler shift measurements can locate emission regions, e.g., *Chandra:*44 i Boo.

**Stellar Flares.** The solar corona magnetic field shows a persistent large-scale structure, but reorganizes itself continuously on small scales. The energy released produces flares spanning 8 decades in energy, and large impulsive reorganizations result in either flares, where energy appears as radiation or in fast particles, or coronal mass ejections (CME), where it appears as plasma kinetic energy. Flares have been observed on many single stars, of class G through M, and on many binaries. They are in some cases orders of magnitude more energetic than solar flares. Observations of a diverse stellar sample can be used to determine flare evolution in various strong lines and hence at various temperatures, and to measure flow velocities, NT line broadening, abundances, and $N_e$ in the stronger flares. The goal is to understand how flares arise on other stars and to interpret results in the light of solar flare research. Dynamical measurements are important as velocities can reach 600 km s$^{-1}$ (AB Dor), comparable to the largest solar flares. Also solar flare X-ray emission lines are strongly broadened (widths 100-300 km s$^{-1}$).

**Abundance Anomalies.** Element abundances in the solar corona differ substantially from photospheric values, which are related to the first ionization potential (FIP): low FIP elements (<10 eV; e.g., Mg, Si, Fe) show enhancements (~x4) when compared to high FIP elements (>10 eV; e.g., O, Ne, Ar). Such anomalies are present also in selected stellar coronae and the full disk solar spectrum. Two satisfactory models have appeared: (i) Ion migration during magnetic flux tube reconnection in the chromosphere, and (ii) Waves in wave-heated loops. Key observations require sensitive high resolution EUV spectroscopy over 170-260 Å, allowing detection of low fluxes of Ar XI-XIV and Fe IX-XIV, in contrast to *EUVE*, where only S was reliably observed.

## IV. Cosmological Implications

In addition to the great discovery potential provided at high *z*, other EUV measurements have cosmological implications, but require an instrument beyond a SMEX. First, some CVs may be progenitors of the Type Ia SNe. The plans to constrain dark energy using large statistical samples of Type Ia SNe should be backed by better understanding of the progenitors. Their final states are the initial conditions for the SN. One progenitor candidate worthy of EUV observation is the double-degenerate binary RX J0806.3+1527 (321.5 sec orbital period).[6] Second, with very high spectral resolving power (R~30,000), the ISM $^3$He and $^4$He components at 304 Å may be separated. The deuterium abundance established by *HST* and *FUSE* places an upper limit on the present baryon density, and the sum of D+$^3$He provides a lower limit. There is some disagreement in $^3$He/$^4$He measurements obtained with other techniques, and models predict that $^3$He/$^4$He should increase in time through chemical evolution. Thus, EUV measurements of the ISM in locations not containing H II regions are extremely important first in establishing baryon density and second in constraining stellar evolution and galactic composition.

## V. Technical, Laboratory, and Theoretical Requirements

Observationally, the EUV has been largely under-utilized with only three satellite instruments flown. The *EUVE* and *ROSAT*-WFC surveys produced catalogs (~1000 targets). However, spectroscopy was limited by the sensitivity (~1 cm$^2$) and resolving power (R~300) of the *EUVE* spectrometer. There are no planned successors to *CHIPS*, which can detect only diffuse emission. Dynamical timescales of stellar coronae and CVs (seconds-to-days) require that next-generation EUV satellite instruments have effective area >20 cm$^2$ and spectrometers have resolving powers 3,000-10,000. Fortunately the technologies of multilayer coatings and ion-etched optics have matured so that normal-incidence EUV instruments with the required performance are practical; no other technology has shown such promise. *J-PEX* has a resolving power >4000 and an effective area of 7 cm$^2$ at 235 Å, beyond the *Chandra* wavelength limit. A proposed SMEX (*APEX*), dedicated to high-resolution EUV spectroscopy, is shown in Figs. 2 and 3.[7-8] Although a single spectrometer has a narrow waveband limited primarily by the multilayer, the suite of spectrometers covers a broad range. Moreover, spectral lines tend to bunch in wavelength, thus spectrometer designs can be optimized.[8] *APEX* would have a resolving power greater than *EUVE* and *Chandra* by factors of 30 and 5, respectively, and an order of magnitude increase in sensitivity.[8] It could accomplish most of the non-cosmological science objectives presented in this White Paper. Sounding rockets can make significant progress on specific objectives, and instruments can be upgraded into a small satellite payload (~3 month mission), which for example could efficiently survey DA WDs for He and heavy elements.[9]

MIDEX or larger instruments can include more challenging galactic and cosmological studies.

Significant laboratory advances are required in the years 2010-2020. Current atomic data are incomplete and inaccurate in wavelength and/or oscillator strengths and opacity, a problem shared with other wavebands (FUV, X-ray) as instruments of higher spectral resolution are flown. WD theoretical models, while enjoying considerable advances during the past decade, must be improved to incorporate stratified atmospheric structures and more accurate mass-loss mechanisms to the ISM. Theoretical comparisons of WDs to NSs classes will prove fruitful. Refinement of stellar coronal models is ongoing, necessary, and with synergy to solar research.

## VI. Summary

Million degree plasmas are ubiquitous in the Universe, and understanding their nature is fundamental to astrophysics. EUV measurements provide a unique, critical, and powerful tool, complementary to those at other wavelengths. With underpinning by a mature instrument technology, there is great opportunity here for exciting discoveries. We strongly recommend that the Astro2010 decadal survey endorse at least a SMEX-class mission and associated research in this often overlooked waveband.

---

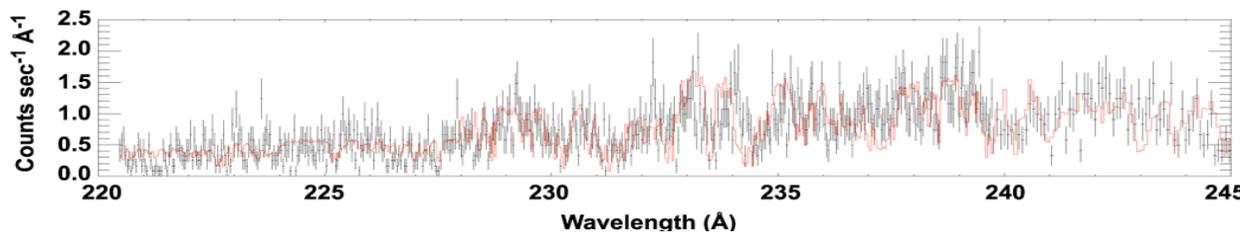

Fig. 1 High-resolution EUV spectrum (black) of the WD G191-B2B, obtained in a 300-second *J-PEX* sounding rocket observation.[1-2] The best fit spectrum (red) yields a photospheric He abundance of $1.6 \times 10^{-6}$ and a LISM He II column density of $5.97 \times 10^{17}$ cm$^{-2}$.

Fig 2. SMEX: 8 EUV spectrometers[7]     Fig. 3. SMEX Effective Area and Resolving Power (R)[7]

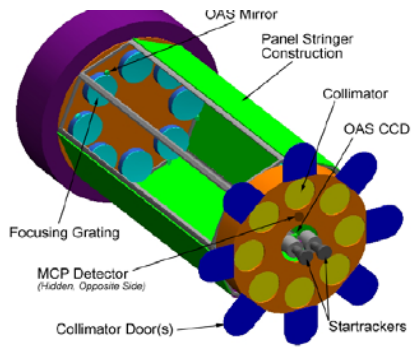
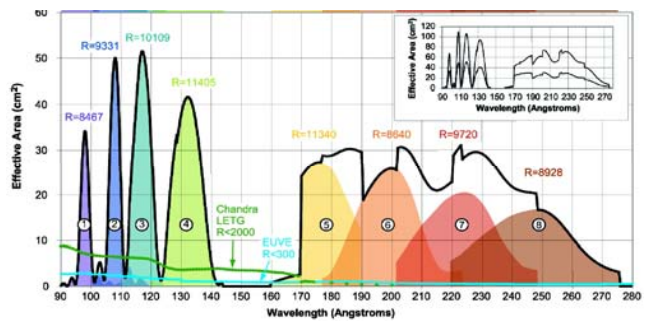

7